# Comment on "Scattering of electromagnetic plane wave by an impedance strip embedded in homogeneous isotropic chiral medium"

Hossein Davoudabadifarahani , Behbod Ghalamkari *

*Computer and Electrical Engineering Department, Science and Research Branch, Islamic Azad University, Tehran, Iran*

**Abstract**

The goal of this paper is to present a previously published work [1] in an errorless form. The work has studied the scattering of electromagnetic plane wave by an impedance strip placed in homogeneous isotropic chiral medium using Kobayashi Potential (KP) method; that has been an important, valuable and attractive investigation in the electromagnetic scattering, especially in KP method. Unfortunately, the study has some basic errors that prevent interesting readers from understanding the investigation. Finally, the formulation of this paper is validated by [2].

**1. Introduction**

One of methods to analyze the electromagnetic scattering problems is Kobayashi Potential (KP); which is a fast and accurate method. KP method analytically satisfies the radiation condition and also boundary conditions are automatically satisfied using Weber-Schafheitlin integrals; which generally, increases the accuracy and decreases the calculation time [3]. These are significant advantages compared with a lot of numerical methods and commercial softwares.

On the other hand, chiral medium plays an important role in the electromagnetic scattering [3, 4]; and has different applications like zero radar cross section [5]. Also, a method for determining the chirality parameter of a chiral medium is proposed in [5], too.

So it is important to understand the interaction between the medium and basic structures such as impedance strip. In the following, the wrong formulations of [1] are corrected; and the equations of this paper are labeled as the same ones of [1].

## 2. Formulation of the problem and KP method

The first errors are the Eqs. (6a) and (6b) which are the scattered Beltrami fields from the strip with unknown weighting functions.

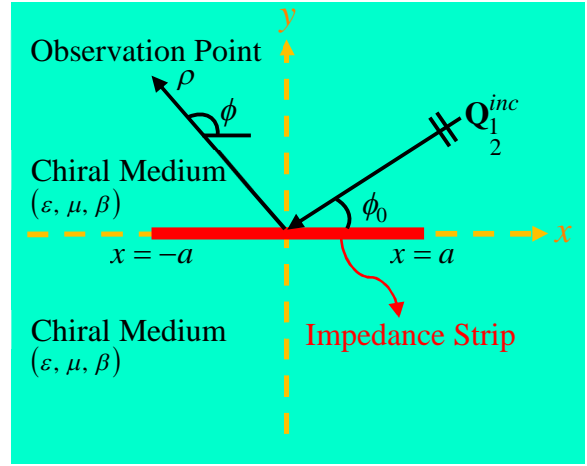

Figure 1. Geometry of the problem

It is known that the Sommerfeld radiation condition should be satisfied by the scattered fields; but there is a disagreement between the condition and the Eqs. (6a) and (6b) of [1]. In other words, the amounts of the scattered fields at infinity ($y_a = \infty$ for Eqs. (6a) and $y_a = -\infty$ for Eqs. (6b)) are equal to infinity. These basic errors cause the other serious errors of the study of [1]. The correct forms of the scattered Beltrami fields are as below:

$$Q_{1z}^{sca} = \int_0^\infty (g_{1c}(\xi)\cos(x_a\xi) + g_{1s}(\xi)\sin(x_a\xi))\, e^{\left(-\sqrt{\xi^2-\gamma_{1a}^2}\right)y_a} d\xi \ , \ y_a > 0 \qquad (6a)$$

$$Q_{1z}^{sca} = \int_0^\infty (h_{1c}(\xi)\cos(x_a\xi) + h_{1s}(\xi)\sin(x_a\xi))\, e^{\left(\sqrt{\xi^2-\gamma_{1a}^2}\right)y_a} d\xi \ , \ y_a < 0 \qquad (6b)$$

The above equations are correct and authentic; because they satisfy the Sommerfeld radiation condition. In other words, these fields at infinity ($y_a = \infty$ for Eqs. (6a) and $y_a = -\infty$ for Eqs. (6b)) have an amount equal to zero. Eqs. (7a)-(7h) of [1] present the boundary conditions of the problem.

Applying the boundary conditions for $|x_a| \leq 1$ where the strip is located (Eqs. (7a)-(7d)) leads to the following equations:

$$\left(1 - \frac{Z_+}{\eta}\sin\phi_0\right)\left[P_1 e^{-j\gamma_{1a}\cos\phi_0 x_a} - j\eta P_2 e^{-j\gamma_{2a}\cos\phi_0 x_a}\right]$$

$$+\int_0^\infty \left(1 + \frac{jZ_+}{\eta\gamma_{1a}}\sqrt{\xi^2 - \gamma_{1a}^2}\right)\left(g_{1c}(\xi)\cos(x_a\xi) + g_{1s}(\xi)\sin(x_a\xi)\right)d\xi \quad (9a)$$

$$-j\eta\int_0^\infty \left(1 + \frac{jZ_+}{\eta\gamma_{2a}}\sqrt{\xi^2 - \gamma_{2a}^2}\right)\left(g_{2c}(\xi)\cos(x_a\xi) + g_{2s}(\xi)\sin(x_a\xi)\right)d\xi = 0$$

$$\left(1 + \frac{Z_-}{\eta}\sin\phi_0\right)\left[P_1 e^{-j\gamma_{1a}\cos\phi_0 x_a} - j\eta P_2 e^{-j\gamma_{2a}\cos\phi_0 x_a}\right]$$

$$+\int_0^\infty \left(1 + \frac{jZ_-}{\eta\gamma_{1a}}\sqrt{\xi^2 - \gamma_{1a}^2}\right)\left(h_{1c}(\xi)\cos(x_a\xi) + h_{1s}(\xi)\sin(x_a\xi)\right)d\xi \quad (9b)$$

$$-j\eta\int_0^\infty \left(1 + \frac{jZ_-}{\eta\gamma_{2a}}\sqrt{\xi^2 - \gamma_{2a}^2}\right)\left(h_{2c}(\xi)\cos(x_a\xi) + h_{2s}(\xi)\sin(x_a\xi)\right)d\xi = 0$$

$$\left(1 - \frac{\eta}{Z_+}\sin\phi_0\right)\left[P_1 e^{-j\gamma_{1a}\cos\phi_0 x_a} + j\eta P_2 e^{-j\gamma_{2a}\cos\phi_0 x_a}\right]$$

$$+\int_0^\infty \left(1 + \frac{j\eta}{Z_+\gamma_{1a}}\sqrt{\xi^2 - \gamma_{1a}^2}\right)\left(g_{1c}(\xi)\cos(x_a\xi) + g_{1s}(\xi)\sin(x_a\xi)\right)d\xi \quad (9c)$$

$$+j\eta\int_0^\infty \left(1 + \frac{j\eta}{Z_+\gamma_{2a}}\sqrt{\xi^2 - \gamma_{2a}^2}\right)\left(g_{2c}(\xi)\cos(x_a\xi) + g_{2s}(\xi)\sin(x_a\xi)\right)d\xi = 0$$

$$\left(1 + \frac{\eta}{Z_-}\sin\phi_0\right)\left[P_1 e^{-j\gamma_{1a}\cos\phi_0 x_a} + j\eta P_2 e^{-j\gamma_{2a}\cos\phi_0 x_a}\right]$$

$$+\int_0^\infty \left(1 + \frac{j\eta}{Z_-\gamma_{1a}}\sqrt{\xi^2 - \gamma_{1a}^2}\right)\left(h_{1c}(\xi)\cos(x_a\xi) + h_{1s}(\xi)\sin(x_a\xi)\right)d\xi \quad (9d)$$

$$+j\eta\int_0^\infty \left(1 + \frac{j\eta}{Z_-\gamma_{2a}}\sqrt{\xi^2 - \gamma_{2a}^2}\right)\left(h_{2c}(\xi)\cos(x_a\xi) + h_{2s}(\xi)\sin(x_a\xi)\right)d\xi = 0$$

In order to satisfy the other boundary conditions, the Weber-Schafheitlin integrals have been used and have resulted in Eqs. (11a)-(11h) of [1]. The system of Eqs. (11a)-(11h) is solved and the eight weighting functions are calculated as below in terms of unknown expansion coefficients of $A_m - I_m$.

$$\begin{pmatrix} g_{1c}(\xi) \\ h_{1c}(\xi) \end{pmatrix} = \pm \frac{1}{4} \sum_{m=0}^{\infty} (A_m + F_m) J_{2m+3/2}(\xi) \xi^{-3/2} + \frac{1}{4\sqrt{\xi^2 - \gamma_{1a}^2}} \sum_{m=0}^{\infty} (C_m + H_m) J_{2m+1/2}(\xi) \xi^{-1/2} \quad (Q1)$$

$$\begin{pmatrix} g_{2c}(\xi) \\ h_{2c}(\xi) \end{pmatrix} = \pm \frac{j}{4\eta} \sum_{m=0}^{\infty} (A_m - F_m) J_{2m+3/2}(\xi) \xi^{-3/2} - \frac{j\gamma_{2a}}{4\eta\gamma_{1a}\sqrt{\xi^2 - \gamma_{2a}^2}} \sum_{m=0}^{\infty} (C_m - H_m) J_{2m+1/2}(\xi) \xi^{-1/2} \quad (Q2)$$

$$\begin{pmatrix} g_{1s}(\xi) \\ h_{1s}(\xi) \end{pmatrix} = \pm \frac{1}{4} \sum_{m=0}^{\infty} (B_m + G_m) J_{2m+5/2}(\xi) \xi^{-3/2} + \frac{1}{4\sqrt{\xi^2 - \gamma_{1a}^2}} \sum_{m=0}^{\infty} (D_m + I_m) J_{2m+3/2}(\xi) \xi^{-1/2} \quad (Q3)$$

$$\begin{pmatrix} g_{2s}(\xi) \\ h_{2s}(\xi) \end{pmatrix} = \pm \frac{j}{4\eta} \sum_{m=0}^{\infty} (B_m - G_m) J_{2m+5/2}(\xi) \xi^{-3/2} - \frac{j\gamma_{2a}}{4\eta\gamma_{1a}\sqrt{\xi^2 - \gamma_{2a}^2}} \sum_{m=0}^{\infty} (D_m - I_m) J_{2m+3/2}(\xi) \xi^{-1/2} \quad (Q4)$$

It is notable that the Eqs. (Q1)-(Q4) have not been presented in [1]; but they are calculated in this paper for a better understanding.

Substituting the weighting functions of Eqs. (Q1)-(Q4) in Eqs. (9), projection in the functional space with elements Jacobi's polynomial $P_n^{\pm 1/2}(x_a^2)$, and assuming that the upper and the lower surface impedances of the strip are equal $(Z_+ = Z_- = Z)$ result in the following matrix equations:

$$\int_0^{\infty} \left( \frac{1}{\xi\sqrt{\xi^2 - \gamma_{1a}^2}} + \frac{j\eta}{Z\gamma_{1a}\xi} \right) \sum_{m=0}^{\infty} C_m J_{2m+1/2}(\xi) J_{2n+1/2}(\xi) \, d\xi$$

$$+ \int_0^{\infty} \left( \frac{1}{\xi\sqrt{\xi^2 - \gamma_{1a}^2}} + \frac{jZ}{\eta\gamma_{1a}\xi} \right) \sum_{m=0}^{\infty} H_m J_{2m+1/2}(\xi) J_{2n+1/2}(\xi) \, d\xi \quad (13a)$$

$$= -4P_1 \frac{J_{2n+1/2}(\gamma_{1a} \cos\phi_0)}{(\gamma_{1a} \cos\phi_0)^{1/2}}, \quad n = 0, 1, 2, \ldots$$

$$\int_0^\infty \left( -\frac{j\eta}{Z\gamma_{1a}\xi} - \frac{\gamma_{2a}}{\gamma_{1a}\xi\sqrt{\xi^2 - \gamma_{2a}^2}} \right) \sum_{m=0}^\infty C_m J_{2m+1/2}(\xi) J_{2n+1/2}(\xi) \, d\xi$$

$$+ \int_0^\infty \left( \frac{\gamma_{2a}}{\gamma_{1a}\xi\sqrt{\xi^2 - \gamma_{2a}^2}} + \frac{jZ}{\eta\gamma_{1a}\xi} \right) \sum_{m=0}^\infty H_m J_{2m+1/2}(\xi) J_{2n+1/2}(\xi) \, d\xi \quad (13b)$$

$$= 4j\eta P_2 \frac{J_{2n+1/2}(\gamma_{2a}\cos\phi_0)}{(\gamma_{2a}\cos\phi_0)^{1/2}}, \, n = 0, 1, 2, \ldots$$

$$\int_0^\infty \left( \frac{1}{\xi\sqrt{\xi^2 - \gamma_{1a}^2}} + \frac{j\eta}{Z\gamma_{1a}\xi} \right) \sum_{m=0}^\infty D_m J_{2m+3/2}(\xi) J_{2n+3/2}(\xi) \, d\xi$$

$$+ \int_0^\infty \left( \frac{1}{\xi\sqrt{\xi^2 - \gamma_{1a}^2}} + \frac{jZ}{\eta\gamma_{1a}\xi} \right) \sum_{m=0}^\infty I_m J_{2m+3/2}(\xi) J_{2n+3/2}(\xi) \, d\xi \quad (13c)$$

$$= 4jP_1 \frac{J_{2n+3/2}(\gamma_{1a}\cos\phi_0)}{(\gamma_{1a}\cos\phi_0)^{1/2}}, \, n = 0, 1, 2, \ldots$$

$$\int_0^\infty \left( -\frac{j\eta}{Z\gamma_{1a}\xi} - \frac{\gamma_{2a}}{\gamma_{1a}\xi\sqrt{\xi^2 - \gamma_{2a}^2}} \right) \sum_{m=0}^\infty D_m J_{2m+3/2}(\xi) J_{2n+3/2}(\xi) \, d\xi$$

$$+ \int_0^\infty \left( \frac{\gamma_{2a}}{\gamma_{1a}\xi\sqrt{\xi^2 - \gamma_{2a}^2}} + \frac{jZ}{\eta\gamma_{1a}\xi} \right) \sum_{m=0}^\infty I_m J_{2m+3/2}(\xi) J_{2n+3/2}(\xi) \, d\xi \quad (13d)$$

$$= 4\eta P_2 \frac{J_{2n+3/2}(\gamma_{2a}\cos\phi_0)}{(\gamma_{2a}\cos\phi_0)^{1/2}}, \, n = 0, 1, 2, \ldots$$

$$\int_0^\infty \left( \frac{\eta}{Z\xi^2} + \frac{j\sqrt{\xi^2 - \gamma_{1a}^2}}{\gamma_{1a}\xi^2} \right) \sum_{m=0}^\infty A_m J_{2m+3/2}(\xi) J_{2n+1/2}(\xi) \, d\xi$$

$$+ \int_0^\infty \left( \frac{Z}{\eta\xi^2} + \frac{j\sqrt{\xi^2 - \gamma_{1a}^2}}{\gamma_{1a}\xi^2} \right) \sum_{m=0}^\infty F_m J_{2m+3/2}(\xi) J_{2n+1/2}(\xi) \, d\xi \quad (13e)$$

$$= 4P_1 \sin\phi_0 \frac{J_{2n+1/2}(\gamma_{1a}\cos\phi_0)}{(\gamma_{1a}\cos\phi_0)^{1/2}}, \, n = 0, 1, 2, \ldots$$

$$\int_0^\infty \left( \frac{\eta}{Z\xi^2} + \frac{j\sqrt{\xi^2 - \gamma_{2a}^2}}{\gamma_{2a}\xi^2} \right) \sum_{m=0}^\infty A_m J_{2m+3/2}(\xi) J_{2n+1/2}(\xi)\, d\xi$$

$$+ \int_0^\infty \left( -\frac{Z}{\eta\xi^2} - \frac{j\sqrt{\xi^2 - \gamma_{2a}^2}}{\gamma_{2a}\xi^2} \right) \sum_{m=0}^\infty F_m J_{2m+3/2}(\xi) J_{2n+1/2}(\xi)\, d\xi \qquad (13f)$$

$$= -4j\eta P_2 \sin\phi_0 \frac{J_{2n+1/2}(\gamma_{2a}\cos\phi_0)}{(\gamma_{2a}\cos\phi_0)^{1/2}},\ n=0,1,2,\ldots$$

$$\int_0^\infty \left( \frac{\eta}{Z\xi^2} + \frac{j\sqrt{\xi^2 - \gamma_{1a}^2}}{\gamma_{1a}\xi^2} \right) \sum_{m=0}^\infty B_m J_{2m+5/2}(\xi) J_{2n+3/2}(\xi)\, d\xi$$

$$+ \int_0^\infty \left( \frac{Z}{\eta\xi^2} + \frac{j\sqrt{\xi^2 - \gamma_{1a}^2}}{\gamma_{1a}\xi^2} \right) \sum_{m=0}^\infty G_m J_{2m+5/2}(\xi) J_{2n+3/2}(\xi)\, d\xi \qquad (13g)$$

$$= -4j P_1 \sin\phi_0 \frac{J_{2n+3/2}(\gamma_{1a}\cos\phi_0)}{(\gamma_{1a}\cos\phi_0)^{1/2}},\ n=0,1,2,\ldots$$

$$\int_0^\infty \left( \frac{\eta}{Z\xi^2} + \frac{j\sqrt{\xi^2 - \gamma_{2a}^2}}{\gamma_{2a}\xi^2} \right) \sum_{m=0}^\infty B_m J_{2m+5/2}(\xi) J_{2n+3/2}(\xi)\, d\xi$$

$$+ \int_0^\infty \left( -\frac{Z}{\eta\xi^2} - \frac{j\sqrt{\xi^2 - \gamma_{2a}^2}}{\gamma_{2a}\xi^2} \right) \sum_{m=0}^\infty G_m J_{2m+5/2}(\xi) J_{2n+3/2}(\xi)\, d\xi \qquad (13h)$$

$$= -4\eta P_2 \sin\phi_0 \frac{J_{2n+3/2}(\gamma_{2a}\cos\phi_0)}{(\gamma_{2a}\cos\phi_0)^{1/2}},\ n=0,1,2,\ldots$$

Solving the matrix integral equations of Eqs. (13) leads to determining the unknown expansion coefficients. Then, the weighting functions are obtained using Eqs. (Q1)-(Q4). After applying the Saddle point method, the longitudinal components of the scattered left and right handed Beltrami fields for $y > 0$ are calculated as:

$$Q_{1z}^{sca} \approx \frac{1}{4}\sqrt{\frac{\pi}{2\gamma_{1a}\rho_a}}e^{j(\gamma_{1a}\rho_a-\frac{\pi}{4})}$$

$$\times \sum_{m=0}^{\infty}\left[\begin{cases}\{A_m J_{2m+1.5}(\gamma_{1a}\cos(\phi))+F_m J_{2m+1.5}(\gamma_{1a}\cos(\phi))\\-j(B_m J_{2m+2.5}(\gamma_{1a}\cos(\phi))+G_m J_{2m+2.5}(\gamma_{1a}\cos(\phi)))\}\tan(\phi)\\-j(C_m J_{2m+0.5}(\gamma_{1a}\cos(\phi))+H_m J_{2m+0.5}(\gamma_{1a}\cos(\phi)))\\-D_m J_{2m+1.5}(\gamma_{1a}\cos(\phi))-I_m J_{2m+1.5}(\gamma_{1a}\cos(\phi))\end{cases}\times(\gamma_{1a}\cos(\phi))^{-(1/2)}\right] \quad (16a)$$

$$Q_{2z}^{sca} \approx \frac{1}{4}\sqrt{\frac{\pi}{2\gamma_{2a}\rho_a}}e^{j(\gamma_{2a}\rho_a-\frac{\pi}{4})}$$

$$\times \sum_{m=0}^{\infty}\left[\frac{j}{\eta}\begin{Bmatrix}A_m J_{2m+1.5}(\gamma_{2a}\cos(\phi))-F_m J_{2m+1.5}(\gamma_{2a}\cos(\phi))\\-j(B_m J_{2m+2.5}(\gamma_{2a}\cos(\phi))-G_m J_{2m+2.5}(\gamma_{2a}\cos(\phi)))\end{Bmatrix}\tan(\phi) \right.$$
$$\left.+\frac{\gamma_{2a}}{\eta\gamma_{1a}}\begin{Bmatrix}-C_m J_{2m+0.5}(\gamma_{2a}\cos(\phi))+H_m J_{2m+0.5}(\gamma_{2a}\cos(\phi))\\+j(D_m J_{2m+1.5}(\gamma_{2a}\cos(\phi))-I_m J_{2m+1.5}(\gamma_{2a}\cos(\phi)))\end{Bmatrix}\times(\gamma_{2a}\cos(\phi))^{-(1/2)}\right] \quad (16b)$$

Here, $\rho_a = \rho/a$, $\rho$ and $\phi$ are radial distance and observation angle, both in cylindrical coordinate, respectively; but these three parameters have not been defined in [1].

### 3. Numerical results and discussion

Using Eqs. (13a)-(13h), the expansion coefficients are calculated and the scattered fields in the far zone are determined by Eqs. (16a)-(16b). The paper [1] has illustrated its results in Figures 1-6. But, since a lot of equations of [1] are incorrect and contain errors, the reported results are not trustworthy. It is notable that the amount of the permittivity and permeability of the chiral medium $(\varepsilon, \mu)$ have not been exactly expressed in the paper [1] for generating the figures; instead, the impedance of the medium $(\eta = \sqrt{\mu/\varepsilon})$ has been given. Also, it is obvious that almost all the equations are nonlinear functions of the permittivity and permeability of the chiral medium, so the figures cannot be regenerated in this investigation.

The presented formulations in this paper are validated using a previously published investigation [2]; for this purpose, the calculated bi-static scattered electric field is compared with the results of

[2]. The equality between the results confirms that the formulations of this paper are completely correct and authentic.

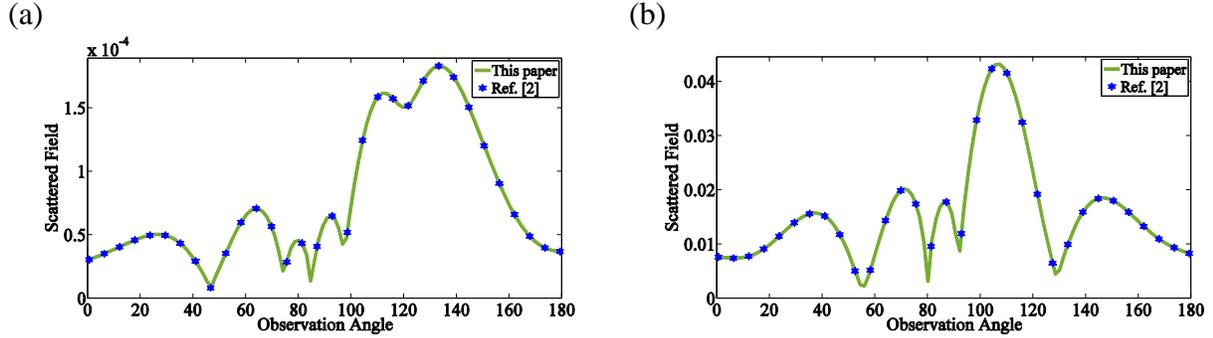

Figure 2. The bi-static scattered electric field over observation angle for $k\beta = 0.2$, $Z \to 0$, $a = 1\lambda$, $\mu = \mu_0$, $\varepsilon = 2\varepsilon_0$ and $\phi_0 = 70°$; The incident field is: (a) left handed Beltrami field, $P_1 = 1$ (b) right handed Beltrami field, $P_2 = 1$.

**Appendix**

The appendix of [1] consists of eight equations where the both surfaces of the strip do not have same impedances ($Z_+$ for the upper surface and $Z_-$ for the lower surface). The correct forms of the equations are calculated as:

$$\sum_{m=0}^{\infty} \int_0^{\infty} \left(1 + \frac{jZ_+\sqrt{\xi^2 - \gamma_{1a}^2}}{\eta \gamma_{1a}}\right) \left\{ \frac{J_{2m+3/2}(\xi)J_{2n+1/2}(\xi)}{\xi^2}(A_m + F_m) \right.$$

$$\left. + \frac{J_{2m+1/2}(\xi)J_{2n+1/2}(\xi)}{\xi\sqrt{\xi^2 - \gamma_{1a}^2}}(C_m + H_m) \right\} d\xi$$

$$+ \int_0^{\infty} \left(1 + \frac{jZ_+\sqrt{\xi^2 - \gamma_{2a}^2}}{\eta \gamma_{2a}}\right) \left\{ \frac{J_{2m+3/2}(\xi)J_{2n+1/2}(\xi)}{\xi^2}(A_m - F_m) \right. \qquad \text{(A1a)}$$

$$\left. - \frac{\gamma_{2a}J_{2m+1/2}(\xi)J_{2n+1/2}(\xi)}{\gamma_{1a}\xi\sqrt{\xi^2 - \gamma_{2a}^2}}(C_m - H_m) \right\} d\xi$$

$$= -4\left(1 - \frac{Z_+\sin\phi_0}{\eta}\right)\left(\frac{P_1 J_{2n+1/2}(\gamma_{1a}\cos\phi_0)}{(\gamma_{1a}\cos\phi_0)^{1/2}} - \frac{j\eta P_2 J_{2n+1/2}(\gamma_{2a}\cos\phi_0)}{(\gamma_{2a}\cos\phi_0)^{1/2}}\right), \; n = 0, 1, 2, \ldots$$

$$\sum_{m=0}^{\infty}\int_0^{\infty}\left(1+\frac{jZ_+\sqrt{\xi^2-\gamma_{1a}^2}}{\eta\gamma_{1a}}\right)\left\{\frac{J_{2m+5/2}(\xi)J_{2n+3/2}(\xi)}{\xi^2}(B_m+G_m)\right.$$

$$\left.+\frac{J_{2m+3/2}(\xi)J_{2n+3/2}(\xi)}{\xi\sqrt{\xi^2-\gamma_{1a}^2}}(D_m+I_m)\right\}d\xi$$

$$+\int_0^{\infty}\left(1+\frac{jZ_+\sqrt{\xi^2-\gamma_{2a}^2}}{\eta\gamma_{2a}}\right)\left\{\frac{J_{2m+5/2}(\xi)J_{2n+3/2}(\xi)}{\xi^2}(B_m-G_m)\right. \quad \text{(A1b)}$$

$$\left.-\frac{\gamma_{2a}J_{2m+3/2}(\xi)J_{2n+3/2}(\xi)}{\gamma_{1a}\xi\sqrt{\xi^2-\gamma_{2a}^2}}(D_m-I_m)\right\}d\xi$$

$$=4\left(1-\frac{Z_+\sin\phi_0}{\eta}\right)\left(\frac{jP_1J_{2n+3/2}(\gamma_{1a}\cos\phi_0)}{(\gamma_{1a}\cos\phi_0)^{1/2}}+\frac{\eta P_2 J_{2n+3/2}(\gamma_{2a}\cos\phi_0)}{(\gamma_{2a}\cos\phi_0)^{1/2}}\right),\ n=0,1,2,\ldots$$

$$\sum_{m=0}^{\infty}\int_0^{\infty}\left(1+\frac{jZ_-\sqrt{\xi^2-\gamma_{1a}^2}}{\eta\gamma_{1a}}\right)\left\{\frac{J_{2m+3/2}(\xi)J_{2n+1/2}(\xi)}{\xi^2}(-A_m-F_m)\right.$$

$$\left.+\frac{J_{2m+1/2}(\xi)J_{2n+1/2}(\xi)}{\xi\sqrt{\xi^2-\gamma_{1a}^2}}(C_m+H_m)\right\}d\xi$$

$$+\int_0^{\infty}\left(1+\frac{jZ_-\sqrt{\xi^2-\gamma_{2a}^2}}{\eta\gamma_{2a}}\right)\left\{\frac{J_{2m+3/2}(\xi)J_{2n+1/2}(\xi)}{\xi^2}(-A_m+F_m)\right. \quad \text{(A1c)}$$

$$\left.-\frac{\gamma_{2a}J_{2m+1/2}(\xi)J_{2n+1/2}(\xi)}{\gamma_{1a}\xi\sqrt{\xi^2-\gamma_{2a}^2}}(C_m-H_m)\right\}d\xi$$

$$=-4\left(1+\frac{Z_-\sin\phi_0}{\eta}\right)\left(\frac{P_1J_{2n+1/2}(\gamma_{1a}\cos\phi_0)}{(\gamma_{1a}\cos\phi_0)^{1/2}}-\frac{j\eta P_2 J_{2n+1/2}(\gamma_{2a}\cos\phi_0)}{(\gamma_{2a}\cos\phi_0)^{1/2}}\right),\ n=0,1,2,\ldots$$

$$\sum_{m=0}^{\infty} \int_0^{\infty} \left(1 + \frac{jZ_-\sqrt{\xi^2 - \gamma_{1a}^2}}{\eta \gamma_{1a}}\right) \left\{ \frac{J_{2m+5/2}(\xi) J_{2n+3/2}(\xi)}{\xi^2} (-B_m - G_m) \right.$$

$$\left. + \frac{J_{2m+3/2}(\xi) J_{2n+3/2}(\xi)}{\xi \sqrt{\xi^2 - \gamma_{1a}^2}} (D_m + I_m) \right\} d\xi$$

$$+ \int_0^{\infty} \left(1 + \frac{jZ_-\sqrt{\xi^2 - \gamma_{2a}^2}}{\eta \gamma_{2a}}\right) \left\{ \frac{J_{2m+5/2}(\xi) J_{2n+3/2}(\xi)}{\xi^2} (-B_m + G_m) \right. \tag{A1d}$$

$$\left. - \frac{\gamma_{2a} J_{2m+3/2}(\xi) J_{2n+3/2}(\xi)}{\gamma_{1a} \xi \sqrt{\xi^2 - \gamma_{2a}^2}} (D_m - I_m) \right\} d\xi$$

$$= 4\left(1 + \frac{Z_- \sin\phi_0}{\eta}\right) \left( \frac{jP_1 J_{2n+3/2}(\gamma_{1a} \cos\phi_0)}{(\gamma_{1a} \cos\phi_0)^{1/2}} + \frac{\eta P_2 J_{2n+3/2}(\gamma_{2a} \cos\phi_0)}{(\gamma_{2a} \cos\phi_0)^{1/2}} \right), \; n = 0, 1, 2, \ldots$$

$$\sum_{m=0}^{\infty} \int_0^{\infty} \left(1 + \frac{j\eta}{Z_+ \gamma_{1a}} \sqrt{\xi^2 - \gamma_{1a}^2}\right) \left\{ \frac{J_{2m+3/2}(\xi) J_{2n+1/2}(\xi)}{\xi^2} (A_m + F_m) \right.$$

$$\left. + \frac{J_{2m+1/2}(\xi) J_{2n+1/2}(\xi)}{\xi \sqrt{\xi^2 - \gamma_{1a}^2}} (C_m + H_m) \right\} d\xi$$

$$+ \int_0^{\infty} \left(1 + \frac{j\eta}{Z_+ \gamma_{2a}} \sqrt{\xi^2 - \gamma_{2a}^2}\right) \left\{ \frac{J_{2m+3/2}(\xi) J_{2n+1/2}(\xi)}{\xi^2} (-A_m + F_m) \right. \tag{A1e}$$

$$\left. + \frac{\gamma_{2a} J_{2m+1/2}(\xi) J_{2n+1/2}(\xi)}{\gamma_{1a} \xi \sqrt{\xi^2 - \gamma_{2a}^2}} (C_m - H_m) \right\} d\xi$$

$$= -4\left(1 - \frac{\eta \sin\phi_0}{Z_+}\right) \left( \frac{P_1 J_{2n+1/2}(\gamma_{1a} \cos\phi_0)}{(\gamma_{1a} \cos\phi_0)^{1/2}} + \frac{j\eta P_2 J_{2n+1/2}(\gamma_{2a} \cos\phi_0)}{(\gamma_{2a} \cos\phi_0)^{1/2}} \right), \; n = 0, 1, 2, \ldots$$

$$\sum_{m=0}^{\infty} \int_0^{\infty} \left(1 + \frac{j\eta}{Z_+\gamma_{1a}}\sqrt{\xi^2 - \gamma_{1a}^2}\right)\left\{\frac{J_{2m+5/2}(\xi)J_{2n+3/2}(\xi)}{\xi^2}(B_m + G_m)\right.$$

$$\left. + \frac{J_{2m+3/2}(\xi)J_{2n+3/2}(\xi)}{\xi\sqrt{\xi^2 - \gamma_{1a}^2}}(D_m + I_m)\right\}d\xi$$

$$+ \int_0^{\infty}\left(1 + \frac{j\eta}{Z_+\gamma_{2a}}\sqrt{\xi^2 - \gamma_{2a}^2}\right)\left\{\frac{J_{2m+5/2}(\xi)J_{2n+3/2}(\xi)}{\xi^2}(-B_m + G_m)\right. \quad\quad\quad \text{(A1f)}$$

$$\left. + \frac{\gamma_{2a}J_{2m+3/2}(\xi)J_{2n+3/2}(\xi)}{\gamma_{1a}\xi\sqrt{\xi^2 - \gamma_{2a}^2}}(D_m - I_m)\right\}d\xi$$

$$= 4\left(1 - \frac{\eta\sin\phi_0}{Z_+}\right)\left(\frac{jP_1 J_{2n+3/2}(\gamma_{1a}\cos\phi_0)}{(\gamma_{1a}\cos\phi_0)^{1/2}} - \frac{\eta P_2 J_{2n+3/2}(\gamma_{2a}\cos\phi_0)}{(\gamma_{2a}\cos\phi_0)^{1/2}}\right), \; n = 0, 1, 2, \ldots$$

$$\sum_{m=0}^{\infty}\int_0^{\infty}\left(1 + \frac{j\eta}{Z_-\gamma_{1a}}\sqrt{\xi^2 - \gamma_{1a}^2}\right)\left\{\frac{J_{2m+3/2}(\xi)J_{2n+1/2}(\xi)}{\xi^2}(-A_m - F_m)\right.$$

$$\left. + \frac{J_{2m+1/2}(\xi)J_{2n+1/2}(\xi)}{\xi\sqrt{\xi^2 - \gamma_{1a}^2}}(C_m + H_m)\right\}d\xi$$

$$+ \int_0^{\infty}\left(1 + \frac{j\eta}{Z_-\gamma_{2a}}\sqrt{\xi^2 - \gamma_{2a}^2}\right)\left\{\frac{J_{2m+3/2}(\xi)J_{2n+1/2}(\xi)}{\xi^2}(A_m - F_m)\right. \quad\quad\quad \text{(A1g)}$$

$$\left. + \frac{\gamma_{2a}J_{2m+1/2}(\xi)J_{2n+1/2}(\xi)}{\gamma_{1a}\xi\sqrt{\xi^2 - \gamma_{2a}^2}}(C_m - H_m)\right\}d\xi$$

$$= -4\left(1 + \frac{\eta\sin\phi_0}{Z_-}\right)\left(\frac{P_1 J_{2n+1/2}(\gamma_{1a}\cos\phi_0)}{(\gamma_{1a}\cos\phi_0)^{1/2}} + \frac{j\eta P_2 J_{2n+1/2}(\gamma_{2a}\cos\phi_0)}{(\gamma_{2a}\cos\phi_0)^{1/2}}\right), \; n = 0, 1, 2, \ldots$$

$$\sum_{m=0}^{\infty} \int_{0}^{\infty} \left(1 + \frac{j\eta}{Z_{-}\gamma_{1a}}\sqrt{\xi^2 - \gamma_{1a}^2}\right) \left\{ \frac{J_{2m+5/2}(\xi)J_{2n+3/2}(\xi)}{\xi^2}(-B_m - G_m) \right.$$

$$\left. + \frac{J_{2m+3/2}(\xi)J_{2n+3/2}(\xi)}{\xi\sqrt{\xi^2 - \gamma_{1a}^2}}(D_m + I_m)\right\}d\xi$$

$$+ \int_{0}^{\infty} \left(1 + \frac{j\eta}{Z_{-}\gamma_{2a}}\sqrt{\xi^2 - \gamma_{2a}^2}\right)\left\{\frac{J_{2m+5/2}(\xi)J_{2n+3/2}(\xi)}{\xi^2}(B_m - G_m)\right. \qquad\qquad (A1h)$$

$$\left. + \frac{\gamma_{2a}J_{2m+3/2}(\xi)J_{2n+3/2}(\xi)}{\gamma_{1a}\xi\sqrt{\xi^2 - \gamma_{2a}^2}}(D_m - I_m)\right\}d\xi$$

$$= 4\left(1 + \frac{\eta\sin\phi_0}{Z_{-}}\right)\left(\frac{jP_1 J_{2n+3/2}(\gamma_{1a}\cos\phi_0)}{(\gamma_{1a}\cos\phi_0)^{1/2}} - \frac{\eta P_2 J_{2n+3/2}(\gamma_{2a}\cos\phi_0)}{(\gamma_{2a}\cos\phi_0)^{1/2}}\right), \; n=0,1,2,\ldots$$